# Unveiling microstructural damage for leakage current degradation in SiC Schottky diode after heavy ions irradiation under 200 V

Xiaoyu Yan, Pengfei Zhai, Chen Yang, Shiwei Zhao, Shuai Nan, Peipei Hu, Teng Zhang, Qiyu Chen, Lijun Xu, Zongzhen Li and Jie Liu

*Abstract*—Single-event burnout and single-event leakage current (SELC) in SiC power devices induced by heavy ions severely limit their space application, and the underlying mechanism is still unclear. One fundamental problem is lack of high-resolution characterization of radiation damage in the irradiated SiC power devices, which is a crucial indicator of the related mechanism. In this letter, high-resolution transmission electron microscopy (TEM) was used to characterize the radiation damage in the 1437.6 MeV $^{181}$Ta-irradiated SiC junction barrier Schottky diode under 200 V. The amorphous radiation damage with about 52 nm in diameter and 121 nm in length at the Schottky metal (Ti)-semiconductor (SiC) interface was observed. More importantly, in the damage site the atomic mixing of Ti, Si, and C was identified by electron energy loss spectroscopy and high-angle annular dark-field scanning TEM. It indicates that the melting of the Ti-SiC interface induced by localized Joule heating is responsible for the amorphization and the formation of titanium silicide, titanium carbide, or ternary phases. These modifications at nanoscale in turn cause the localized degradation of the Schottky contact, resulting in the permanent increase in leakage current. This experimental study provides very valuable clues to thorough understanding of the SELC mechanism in SiC diode.

This work was financially supported by National Key Research and Development Program of China (Grant No. 2022YFB3604001), National Natural Science Foundation of China (Grant Nos. 12035019, 12075290, 62234013), the Youth Innovation Promotion Association of Chinese Academy of Sciences (Grant No. 2020412), the Innovation Center of Radiation Application (No. KFZC2022020601）(Corresponding authors: Pengfei Zhai and Jie Liu)

X. Yan, P. Zhai, C. Yang, S. Zhao, P. Hu, Q. Chen, L. Xu, Z. Li and J. Liu are with the Institute of Modern Physics, Chinese Academy of Sciences, Lanzhou 730000, China and also with the School of Nuclear Science and Technology, University of Chinese Academy of Sciences, Beijing 100049, China (e-mail: zhaipengfei@impcas.ac.cn; j.liu@impcas.ac.cn)
S. Nan is with the Songshan Lake Materials Laboratory, Dongguan 523808, Guangdong, China and also with the Institute of Physics, Chinese Academy of Sciences, Beijing 100190, China
T. Zhang is with the Nanjing Electronic Devices Institute, Nanjing 210016, China

## I. Introduction

SILICON carbide (SiC) power devices have inherent advantages such as high breakdown electric field, high power density and high operating temperature, which were expected to be a promising candidate for space application in the future. However, these devices are extremely susceptible to high-energy heavy ions irradiation [1-9]. The catastrophic single-event burnout (SEB) and single-event leakage current (SELC) in SiC power devices caused by low linear energy transfer (LET) of heavy ions severely limit the potential space application.

It is crucial to explore the underlying mechanism, which is the foundation for the development of radiation-hardened SiC power devices. Therefore, there are numerous experimental and simulation studies dealing with the radiation effects of SEB and SELC in SiC power devices [1-17]. Material damage is widely presumed to be the origin of permanent SELC degradation in SiC power device [1-4,8-10]. However, the nature of material damage responsible for SELC degradation is still under debate [1,10,18-20]. In 2006, Kuboyama et al. [1] reported one low-magnification transmission electron microscopy (TEM) image of radiation damage in SiC Schottky barrier diode (SBD). The observed radiation damage caused by 394 MeV $^{129}$Xe (LET is 73.1 MeV·cm$^2$·mg$^{-1}$) under reverse bias of 160 V is located at the metal contact with the diameter of ~50 nm and length of ~70 nm [1]. This result was cited as the proof supporting that radiation damage originates from a local Joule heating [2,3,9]. The molecular dynamics (MD) simulation also suggested that nanoscale amorphization of SiC can be formed in about 100 ps due to Joule heating, when the SiC SBD biased at 100 V or above is irradiated with single Xe ion (LET is 62.4 MeV·cm$^2$·mg$^{-1}$) [10]. The formation of extended defects in SiC, such as stacking faults, was also suggested to be the origin of SELC degradation [18-20]. Recently, one new deep level transient spectroscopy (DLTS) peak with an activation energy of 0.17 eV below the conduction band edge (Ec) was observed solely in the SiC SBD with SELC signature [20]. Although this energy level should be assigned to Ti peak according to the references [21-23], the origin of this peak was finally assigned to single-layer Shockley-type stacking faults (the energy level is Ec-0.213 eV), because the authors stated that this Ti 'impurity' only appears in the device with SELC signature [20].

To solve this fundamental and long-standing problem, high-resolution and direct characterization of radiation damage in SELC degraded SiC device is crucial.

In this work, high-resolution TEM combined with electron energy loss spectroscopy (EELS) were used to provide a comprehensive information of the radiation damage in SiC Schottky diode induced by 1437.6 MeV $^{181}$Ta ions under 200V.

## II. EXPERIMENTAL DETAILS

The SiC junction barrier Schottky (JBS) diodes were provided by the Nanjing Electronic Devices Institute. The device type was WS3A015120D rated at 1200 V and 15 A. Prior to heavy ions irradiation, the packaging material was removed. The total thickness consists of 4 μm Al, 100 nm Ti, and 10.5 μm SiC epitaxial layer including ~0.5-1 μm buffer layer (varied doping concentration) between the substrate and epitaxial layer. The schematic of the device is shown in Fig. 1(a).

$^{181}$Ta ions irradiation was performed at the Heavy Ion Research Facility in Lanzhou (HIRFL) in the Institute of Modern Physics, Chinese Academy of Sciences. The initial kinetic energy of $^{181}$Ta ions is 2896 MeV. After penetrating through secondary electron detector (7.2 μm Al), Titanium window (14.7 μm), polyethylene terephthalate (PET) membrane (12×2 μm), and 50 mm Air, the residual energy of the $^{181}$Ta ion is 1437.6 MeV. The LET on the surface of the device is 84.7 MeV·cm$^2$·mg$^{-1}$ and the projected range in SiC is 57.6 μm, as calculated by SRIM 2013 code [24]. The projected range is long enough to penetrate through the epitaxial layer, so the insufficient range effect reported by Sengupta et al. [5] can be avoided. The variation of LET in the whole epitaxial layer is quite smooth. The ion flux was about 5.9×10$^5$ ions·cm$^{-2}$·s$^{-1}$. PET membrane was used as the solid-state nuclear track detector to provide the ion fluence as 1.23×10$^9$ ions·cm$^{-2}$ with the uncertainty less than 10%. This ion fluence corresponds to ~12 ions for each 1 μm$^2$, ensuring that the radiation damage can be found easily by TEM. The heavy ions irradiation was performed under normal incidence at room temperature in air conditions.

During the heavy ions irradiation, the device under test (DUT) was biased at the reverse voltage ($V_R$) of 200 V and the leakage current ($I_R$) was monitored by the Keithley 2657A. The measured SEB threshold for this device is about 450 V. It is consistent with the data summarized by Ball et al. [25]. The 200 V reverse bias make sure that the DUT operates in the region of single-event leakage current [1,3,25].

After heavy ion irradiation and post-irradiation electrical characterization of the DUT, a dual Cs-corrected transmission electron microscope (JEM-ARM300F, JEOL Ltd.) operating at 300 kV was used to characterize the damage. The TEM samples were prepared using focused ion beam technique, and the possible element redistribution at the damage site was measured using EELS (GIF Continuum ER 1065, Gatan Inc.).

## III. RESULTS AND DISCUSSION

The leakage current of the DUT during the 1437.6 MeV Ta ions irradiation under the reverse voltage of 200 V is shown in Fig. 1(b). The leakage current increased significantly once the heavy ions beam was on. The current limit was set to be 120 mA to avoid the device complete failure. To reduce the self-heating effect in the SELC degraded device, discontinuous irradiation was adopted in the experiment. In this way, the DUT could cool down naturally during the beam and bias off (~2-5 minutes). In the second and subsequent runs of irradiation, we paused the irradiation intentionally when the leakage current reach about 11 mA to save the cooling down time. Seven runs of irradiation were conducted in total, and the accumulative ion fluence was up to 1.23×10$^9$ ions·cm$^{-2}$. After irradiation the leakage current of the DUT was about 3.3 mA ($V_R$=200 V). Fig. 1 (c) and (d) show the forward and reverse current-voltage (I-V) characteristics of the DUT compared with the three pristine devices under different bias voltages. The post-irradiation electrical characterization was performed 30 days after the end of irradiation. The forward current of these devices showed no obvious change. However, the DUT showed the remarkable reverse leakage current degradation, and the reverse current under 200 V was about 3.9 mA. It indicates that the permanent damage has been induced in the DUT when it was irradiated under 200 V, and the damage cannot be recovered after 30 days storage at room temperature. As a comparison, the reverse leakage current of the same type device with SEB signature is estimated to be about 1 mA, 48 mA and 193 mA if biased at 1V, 50 V, and 200 V, respectively, according to the reverse resistance of the SEB device is measured to be only 1036 Ω. It confirms that the irradiated DUT with 200 V bias is SELC degraded.

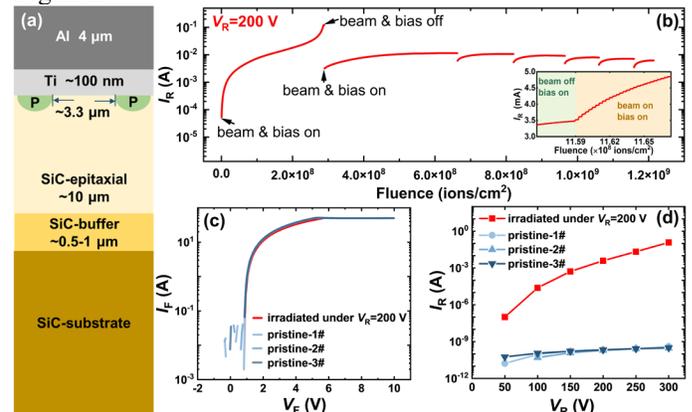

Fig. 1. The schematic of the DUT (a, not to scale). The curves of leakage current during the heavy ions irradiation under the $V_R$ of 200 V (b). The forward (c) and reverse (d) I-V curves of the SiC JBS diodes before and after irradiation under different bias voltages.

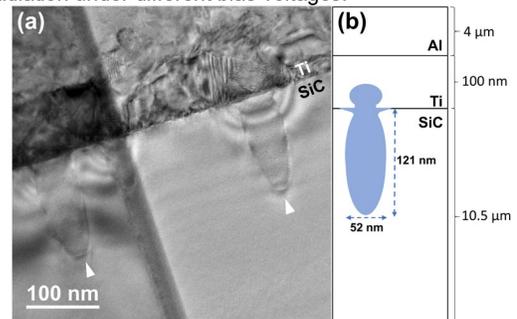

Fig. 2. The TEM image (a) and the schematic (b, not to scale) of the typical damage in the interface of the Schottky metal (Ti) and epitaxial layer (SiC) after irradiation under the $V_R$ of 200 V.

Two typical damage sites in the DUT characterized by TEM are presented in Fig. 2. They are located at the interface of the Schottky metal (Ti) and epitaxial layer (SiC). This location is exactly the one of the most possible damage sites predicted by the previous technology computer aided design (TCAD) simulations [2,3,9]. It is also consistent with the experimental result by Kuboyama et al. [1]. Forty-eight damage sites in SiC side were measured, and the length and diameter of the damage site are 121 ± 25 nm and 52 ± 7 nm, respectively. The uncertainty is from the standard deviation. The measured length of the damage is much larger than that (70 nm) in the irradiated SiC SBD with 394 MeV $^{129}$Xe (LET is 73.1 MeV·cm$^2$·mg$^{-1}$), under the $V_R$ of 160 V [1]. In contrast, the diameter increases slightly (52 nm vs. 50 nm [1]). The schematic diagram of the damage is given in Fig. 2(b). Due to the Abbe diffraction limit, the position accuracy of the emission microscope (EMMI) is only about 0.3 μm. It is difficult to position the SELC failure site of ~50 nm in diameter using this method. It can explain reasonably why Peng et al. [26] did not observe any radiation damage before.

Interestingly, none of the 48 damage sites were located at the interface of the metal Ti and p region. TCAD simulation qualitatively confirmed that the localized electric field here is significantly moderate [3,27]. We have also tried to find the possible damage site at the interface of the epitaxial layer and substrate of SiC, because it is the other location where the electric field is highly intense predicted by some TCAD simulations [2,3,27]. However, no obvious structural change was identified. It indicates that the buffer layer designed in the DUT can mitigate the electric field peak in this interface [28,29].

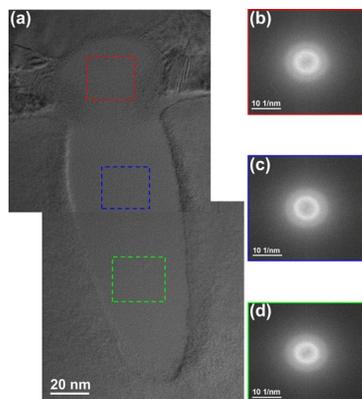

Fig. 3. The high-resolution TEM image (a) of the typical damage in the DUT under $V_R$ of 200 V. (b)-(d) are the corresponding FFT pattern of the regions marked in (a).

High-resolution TEM combined with the fast Fourier transforms (FFT) analysis could provide a comprehensive information about the individual damage as follows (Fig. 3): an amorphous hillock with two little amorphous arms on the interface are observed, while beneath the hillock, there are a long amorphous body in SiC. Moreover, the interface here is no longer sharp. It should be noted that only point defects will be formed, even if the SiC material is solely irradiated by 2640 MeV $^{238}$U ions with the LET of 114.9 MeV·cm$^2$·mg$^{-1}$ [30]. It can be concluded that the observed amorphous damage is caused by heavy ion irradiation and highly localized electric field. This direct and high-resolution TEM image of the radiation damage supports that the MD simulation results by Javanainen et al. [10].

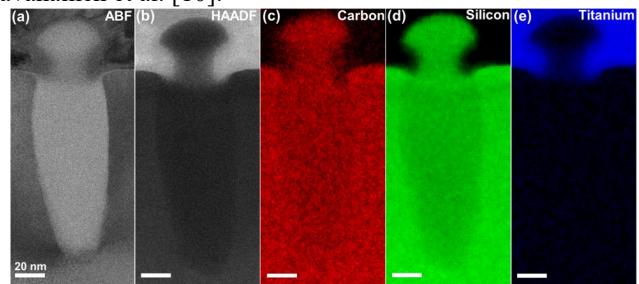

Fig. 4. The ABF-STEM image (a), HAADF-STEM image (b) and EELS elements mapping (c-e) of the typical damage in the DUT under $V_R$ of 200 V.

The annular bright field scanning TEM (ABF-STEM), high-angle annular dark-field STEM (HAADF-STEM) [31] and the EELS mapping analysis further identified the element redistribution in the damage site for the first time (Fig. 4). It shows that the elements C and Si flowed into the Ti layer, and the profile coincides well with the amorphous hillock, while the element Ti mainly flowed into the near-surface of SiC corresponding to the two amorphous arms (Fig. 3 and 4). It indicates that localized melting occurred and induced amorphization and prominent atomic mixing. This atomic mixing forming titanium silicide, titanium carbide, or ternary phases, and the amorphization of SiC could locally lower the Schottky barrier height [3,8,10,32]. This degradation of Schottky contact at nanoscale could act as the permanent leakage path for the reverse leakage current degradation. Interestingly, according to our work, it indicates that the appearance of the new DLTS peak ($E_C$-0.17 eV) only in SiC SBD with SELC signature [20] is probably due to the atomic mixing in the region of radiation damage. Therefore, the Ti 'impurity' was detected only in the SELC degraded device [20].

The above experimental results cannot be simulated solely by TCAD, because the atomic transport, such as amorphization, recrystallization, and atomic mixing, is not involved. TCAD combined with the molecular dynamics simulation is likely to give a comprehensive understanding in the future work [10], and the interface of the Schottky metal and SiC should be considered [11].

## IV. Conclusion

The detailed information of the damage in the irradiated SiC JBS diode with high-energy heavy ions under 200 V are presented. For the first time, the atomic mixing of Ti, C and Si in the damage site was identified. These results provide a new insight into the mechanism of SELC in SiC Schottky diode. It indicates that modulating the electric field distribution near the Schottky contact [15] and choosing a refractory Schottky metal should be a promising alternative to enhance the tolerance to the SELC and SEB.

## Acknowledgment

We would like to thank the accelerator staff of the HIRFL for providing the high-quality heavy ion beam.